# Realization of a three-dimensional photonic topological insulator

*Yihao Yang[1,2,3,4], Zhen Gao[3,4,\*], Haoran Xue[3,4], Li Zhang[1,2], Mengjia He[1,2], Zhaoju Yang[3,4], Ranjan Singh[3,4], Yidong Chong[3,4], Baile Zhang[3,4,\*] and Hongsheng Chen[1,2,\*]*

[1]State Key Laboratory of Modern Optical Instrumentation, The Electromagnetics Academy at Zhejiang University, Zhejiang University, Hangzhou 310027, China.

[2]Key Laboratory of Micro-Nano Electronics and Smart System of Zhejiang Province, College of Information Science and Electronic Engineering, Zhejiang University, Hangzhou 310027, China.

[3]Division of Physics and Applied Physics, School of Physical and Mathematical Sciences, Nanyang Technological University, 21 Nanyang Link, Singapore 637371, Singapore.

[4]Centre for Disruptive Photonic Technologies, The Photonics Institute, Nanyang Technological University, 50 Nanyang Avenue, Singapore 639798, Singapore.

\*Correspondence to: gaozhen@ntu.edu.sg, blzhang@ntu.edu.sg, hansomchen@zju.edu.cn

**Confining photons in a finite volume is in high demand in modern photonic devices. This motivated decades ago the invention of photonic crystals, featured with a photonic bandgap forbidding light propagation in all directions [1-3]. Recently, inspired by the discoveries of topological insulators (TIs) [4,5], the confinement of photons with topological protection has been demonstrated in two-dimensional (2D) photonic structures known as photonic TIs [6-8], with promising applications in topological lasers [9,10] and robust optical delay lines [11]. However, a fully three-dimensional (3D) topological photonic bandgap has never before been achieved. Here, we experimentally demonstrate a 3D photonic TI with an extremely wide (> 25% bandwidth) 3D topological bandgap. The sample consists of split-ring resonators (SRRs) with strong magneto-electric coupling and behaves as a "weak TI", or a stack of 2D quantum spin Hall insulators. Using direct field measurements, we map out both the gapped bulk bandstructure and the Dirac-like dispersion of the photonic surface states, and demonstrate robust photonic propagation along a non-planar surface. Our work extends the family of 3D TIs from fermions to bosons and paves the way for applications in topological photonic cavities, circuits, and lasers in 3D geometries.**

Photonic bandgap materials, also known as photonic crystals, are engineered materials capable of confining photons due to photonic bandgaps that forbid the propagation of electromagnetic waves (i.e., light waves governed by the classical Maxwell equations) in all directions. While electronic bandgaps are a long-established concept, it was only in the late 1980s that photonic bandgap materials were theoretically proposed as an electromagnetic analog of semiconductor crystals [1,2]. They were then experimentally realized in the form of a 3D photonic crystal with a complete bandgap at microwave frequencies [3]. Subsequent researchers have realized 3D photonic crystals at optical frequencies [12,13], and shown that photons can be confined in cavities or optical circuits by embedding point, line, or volume defects in a 3D photonic crystal [13]. The photon-confining capability of a photonic crystal is generally determined by the width of its bandgap.

During the past two decades, condensed matter physics has been revolutionized by the introduction of topological classifications of phases of matter, including 2D and 3D TIs [4,5]. While 2D TIs host topologically-protected one-way *edge* states, 3D TIs exhibit topological *surface* states; these surface states are not unidirectional, but behave as 2D massless Dirac fermions [4,5]. Analogues of 2D TIs, based on numerous different design principles, have been implemented in photonics [14,15], and can be used to implement novel topologically protected lasers [9,10] and optical delay lines [11]. However, in these 2D systems, photonic confinement in the third (out-of-plane) direction is achieved by non-topological means such as index guiding. A 3D topological photonic bandgap, which can achieve topological confinement of photons in all three spatial directions, has never previously been achieved. Although topologically nontrivial 3D bandstructures have been demonstrated in Weyl photonic crystals [16-18], these are *ungapped* systems that cannot be used to confine light.

Recently, there have been several theoretical proposals for realizing a 3D topological photonic bandgap [19-23]. One uses high-index magneto-optic materials to generate a bandstructure analogous to a "strong" TI (which has an odd number of surface Dirac cones), albeit one with an incomplete bandgap; this has, however, proven challenging to fabricate [20]. Another recent proposal involves a photonic "weak" TI, which possesses an even number of surface Dirac cones [21]. Weak TIs emerge from stacking layers of 2D quantum spin Hall insulators with appropriate interlayer couplings [24]. Although weak-TI surface states were originally considered to be unprotected against disorder, recent research has revealed that they are robust against disorder, so long as time-reversal symmetry and the existence of the bandgap are maintained [25,26].

Here, we report on the realization of a 3D photonic TI, featuring a complete and extremely wide topological bandgap. We have experimentally mapped out the bulk bandstructure and the dispersion of the surface states along an internal domain wall. We show explicitly that the surface states have the predicted form of a Dirac cone [21], which is the key distinguishing feature of a 3D photonic TI. We also experimentally demonstrate robust photonic transport in 3D along a sharply twisted internal domain wall. Unlike the proposal of Slobozhanyuk et al. [21], the present design utilizes a 3D array of metallic SRRs, which are classical electromagnetic artificial "atoms" that serve as building blocks for metamaterials [27]. The resonance-enhanced bianisotropy of the SRRs plays a role analogous to the strong spin-orbit coupling in TI materials [4,5], allowing for a topological bandgap with width > 25%. This exceeds even previously-demonstrated topological bandgap widths in 2D, which have been 10% or less [7,9-11,14,28,29], and substantially exceeds the gap widths in previous 3D proposals, which were on the order of a few percents or incomplete [19-23].

We start with a photonic crystal design featuring an ungapped bandstructure with 3D Dirac points. As depicted in Fig. 1a, the photonic crystal has a unit cell consisting of six connected metallic SRRs. The crystal is formed by arranging the unit cells in a triangular lattice in the *x-y* plane, as shown in Fig. 1b, and stacking identical layers along the *z*-direction. The background material is Teflon woven glass fabric copper-clad laminate, with relative permittivity 2.5. Note that the back-to-back arrangement of the SRRs cancels the bianisotropy at the K and K′ points in the Brillouin zone [30]. The lattice has a mirror ($z \to -z$) symmetry, which we denote by $\sigma_z$. For fine-tuned lattice parameters, the photonic bandstructure exhibits frequency-isolated 3D Dirac points [21] (i.e., doubly-degenerate Weyl points), with four-fold degeneracy at the band-crossing points at K and K′, as shown in Fig. 1d.

The design for a wide-gap photonic TI is shown in Fig. 1c. These SRRs are *not* arranged back-to-back and $\sigma_z$ is broken. This unit cell can be formed by removing the upper three SRRs in the unit cell of Fig. 1a and adjusting the *z* periodicity accordingly. The resultant bandstructure is shown in Fig. 1e, and exhibits a wide (> 25%) bandgap. Although the relationship between the two photonic crystal designs is not immediately apparent, we show in the Methods that the bandgap in Fig. 1e is continuously deformable into the infinitesimal bandgap opened at the Dirac points when the $\sigma_z$ symmetry of Fig. 1a is perturbatively broken. In the original $\sigma_z$-symmetric structure, the electric and magnetic dipoles have opposite parity and form decoupled modes under in-plane propagation ($k_z$=0). Breaking $\sigma_z$ induces a bianisotropic coupling between the in-plane electric and magnetic dipoles, generating two hybrid modes in each of the lower and upper bands. Near the K and K' valleys, the pair of eigenmodes within each band have electric and magnetic components that are respectively in-phase or out-of-phase, corresponding to pseudo-spin-up or pseudo-spin-down, for both lower and upper bands [15,21,29,31]. This form of bianisotropic coupling has previously been shown to play a role analogous to the Kane-Mele spin-orbit coupling in quantum spin Hall materials [4,5,15,21,31]. See Methods for theoretical characterization of the topological properties.

The experimental sample of photonic TI is shown in Fig. 2a (see Methods). In the first set of experiments, illustrated in Fig. 2b, a dipole source antenna is inserted from the bottom center of the *y-z* plane, indicated by the red (blue) dotted line, 4 unit cells into the bulk; a second dipole antenna, acting as the probe, is inserted inside the sample from the hole marked with a red (blue) dot in Fig. 2a, at a depth of 8 unit cells into the bulk, to measure the transmission of the surface (bulk) states. We wrapped all sides of the sample with microwave absorbers. The frequency dependence of the transmittance, along the domain wall, and in bulk, is shown in Fig. 2c. We observe an approximately 20 dB drop in the transmittance of the bulk states, extending from approximately 4.3 GHz to 6.0 GHz, corresponding to a bulk bandgap. Along the domain wall, however, the transmittance remains high throughout the frequency range, indicating the existence of surface states. Numerical simulations reveal that pseudospin-momentum locking occurs along the isofrequency contours of the surface Dirac cone, which is another key feature of 3D photonic TIs [21,31] (see Methods).

To probe the bulk bandstructure, we measure the electromagnetic response in the *y-z* plane (indicated by the green dotted line in Fig. 2a) away from the domain wall. The probe dipole antenna is fixed to a robotic arm, and we map the complex electric field patterns in the selected plane. After Fourier transformation to reciprocal space, we obtain the plot shown in Fig. 2d, which closely matches the numerically-computed projected bandstructure shown in Fig. 2e. Note that there is a complete photonic bandgap in the bulk, with almost the same frequency range of the bandgap in Fig. 2c.

Next, we repeat the field measurements with the source and probe located along the 2D domain wall. The surface bandstructure, shown in Fig. 3a, reveals a family of surface states that span the frequency range of the bulk bandgap, forming a conical dispersion curve. The surface Dirac point occurs at 5.1 GHz, near the mid-point of the bulk bandgap, and along the Γ-X high-symmetry line (near the projection of the K point of the 3D Brillouin zone). A second Dirac point near K′ is not shown in this plot. The isofrequency plots in the 2D reciprocal space confirm that the dispersion is indeed conical, as shown in the insets of Fig. 3a. These results closely match the numerically computed surface bandstructure, as shown in Fig. 3b. By placing the probes in adjacent unit cells, away from the domain wall, we verify that the surface states are tightly confined to the domain wall with a penetration depth of around 10.3 mm (see Methods).

Finally, we demonstrate robust propagation of the surface states along a non-planar domain wall. As schematically shown in Fig. 4a, the wall is sharply twisted, with two 60° corners. A waveguide (marked by a black arrow) launches topological surface states that propagate to the right. The measured transmission along the twisted domain wall (at the position of the red dot) is comparable to that measured along the previous straight domain wall of the same length, as shown in Fig. 4b. In comparison, the measured transmission in the bulk (at the green dot in Fig. 4a) is substantially lower, due to the bulk bandgap. We also performed a 3D map of the field distributions at 4.7 GHz (see Methods). The results, shown in Fig. 4c, demonstrate that the topological surface states flow robustly along the surface, including around the two sharp corners. Extracting different $k_z$ components, we plot in Figs. 4d-f the field distributions with $k_z = 0$ /m, $k_z = 31.4$ /m, and $k_z = 62.8$ /m, respectively. (These are typical values in the range of the allowable $k_z$; see Methods). Negative values of $k_z$ behave similarly and are therefore omitted. The results match well with the numerical results (see Methods).

Our work thus demonstrates a classical photonic analog of a 3D TI. The realization of a 3D topological photonic bandgap opens the door to a wide range of novel topological photonic devices, such as topological photonic lasers [9,10] and circuits in previously inaccessible 3D geometries. This also provides the possibility to study topological quantum optics beyond 2D [28] in new aspects such as spontaneous emission [1] in a fully 3D topological cavity. Although our demonstration has been carried out at microwave frequencies, these design principles should be generalizable to other frequency regimes. Current advances in 3D SRR fabrication have shown the feasibility of operating frequencies at terahertz and infrared frequencies [32]. An implementation based on all-dielectric metamaterials [21] may allow for a 3D photonic TI in the optical regime. The current study has focused on a photonic crystal for electromagnetic waves, but a similar lattice design may be applied to other bosonic systems, such as acoustic and mechanical structures [33].

## Acknowledgments

We thank Q. Yan at Zhejiang University, L. Lu at the Chinese Academy of Sciences, and J. C. W. Song at Nanyang Technological University for helpful discussions. The work at Zhejiang University was sponsored by the National Natural Science Foundation of China under Grant Nos. 61625502, 61574127, 61601408, 61775193 and 11704332, the ZJNSF under Grant No. LY17F010008, the Top-Notch Young Talents Program of China, the Fundamental Research Funds for the Central Universities, and the Innovation Joint Research Center for Cyber-Physical-Society System. Y. D. Chong and B. Zhang acknowledge the support of Singapore Ministry of Education under Grants No. MOE2015-T2-1-070, MOE2015-T2-2-008, MOE2016-T3-1-006 and Tier 1 RG174/16 (S). Y. Yang and R. Singh acknowledge the support of Singapore Ministry of Education under Grants No. MOE2015-T2-2-103.

## Authors Contributions

Y.Y. initiated the original idea. Y.Y., B.Z., and H.C. designed the experiment. Y.Y., Z.G., M.H., and L.Z. fabricated samples. Y.Y. and Z.G. carried out the measurement and analyzed data. Y.Y., H.X., L.Z., and Z.Y. performed simulations. Y.Y., H.X., B.Z., M.H., Z.Y., H.C., and Y.C. provided the theoretical explanations. R.S. assisted in part of the experiment. Y.Y., Y.C., B.Z., and H.C. supervised the project.

## Author Information

The authors declare no competing financial interests. Reprints and permissions information is available at www.nature.com/reprints. Readers are welcome to comment on the online version of the paper. Correspondence and requests for materials should be addressed to Z. G. (gaozhen@ntu.edu.sg), B. Z. (blzhang@ntu.edu.sg), or H. C. (hansomchen@zju.edu.cn).

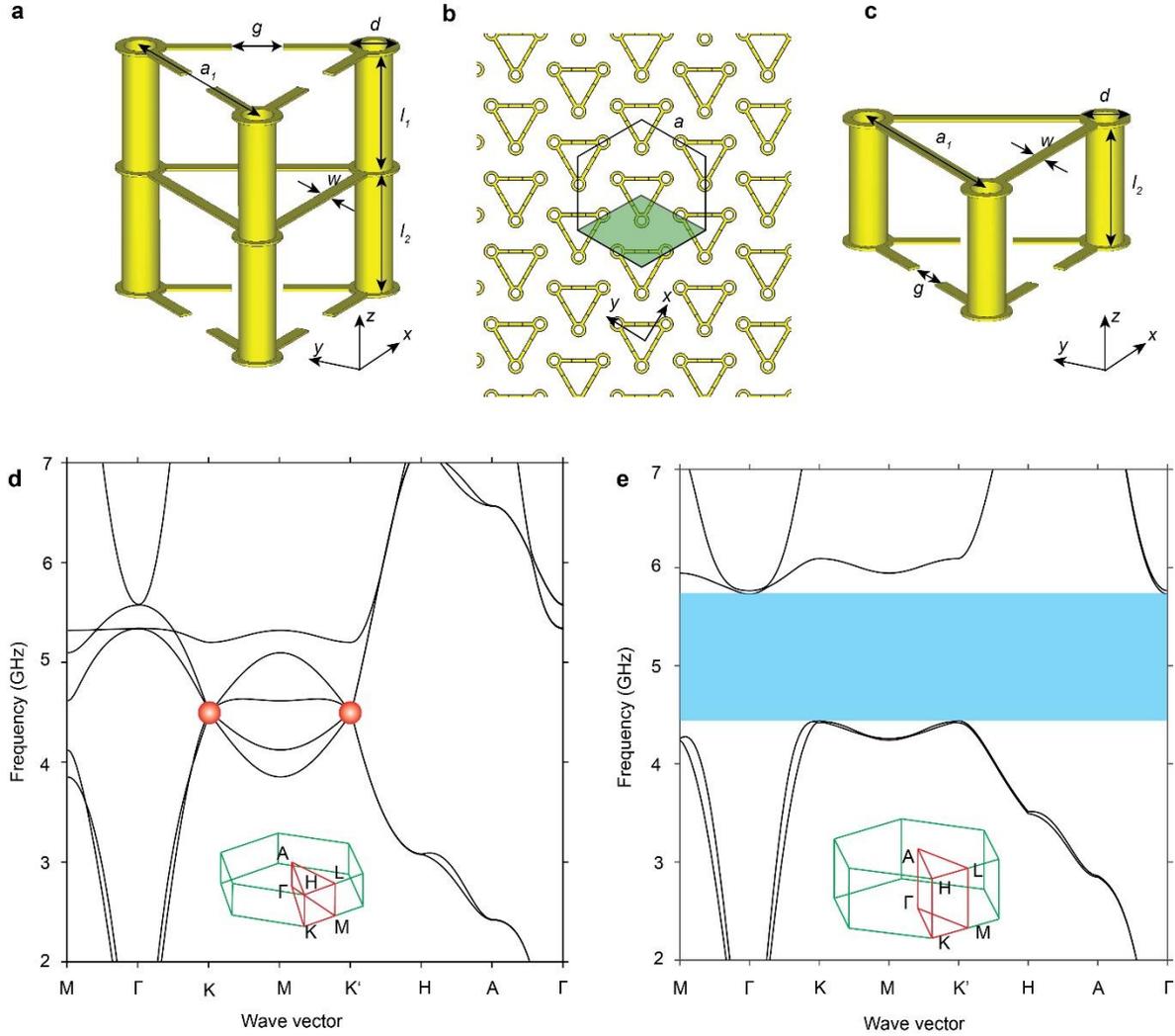

**Figure 1. Design of photonic structures with 3D Dirac points and 3D topological bandgap.** (a-b) Unit cell and top view of a hexagonal lattice with 3D Dirac points. The structure parameters are $l_1=l_2=5$ mm, $g=2$ mm, $a_1=9$ mm, $w=0.5$ mm, $d=1.5$ mm, and lattice constant $a=10.4$ mm. The periodicity in the $z$-direction is 11.3 mm. The background dielectric material has a relative permittivity of 2.5. The green rhombus represents a unit cell. (c) Unit cell giving rise to a 3D topological bandgap. The top three SRRs in (a) are removed. The three remaining SRRs have the same geometrical parameters as in (a). The periodicity in the $z$-direction is 5.65 mm. (d-e) Band diagrams of the crystals with unit cells shown in (a) and (c), respectively. Insets show the 3D Brillouin zones for each unit cell. The red dots in (d) are the 3D Dirac points. The blue region in (e) represents the complete topological bandgap.

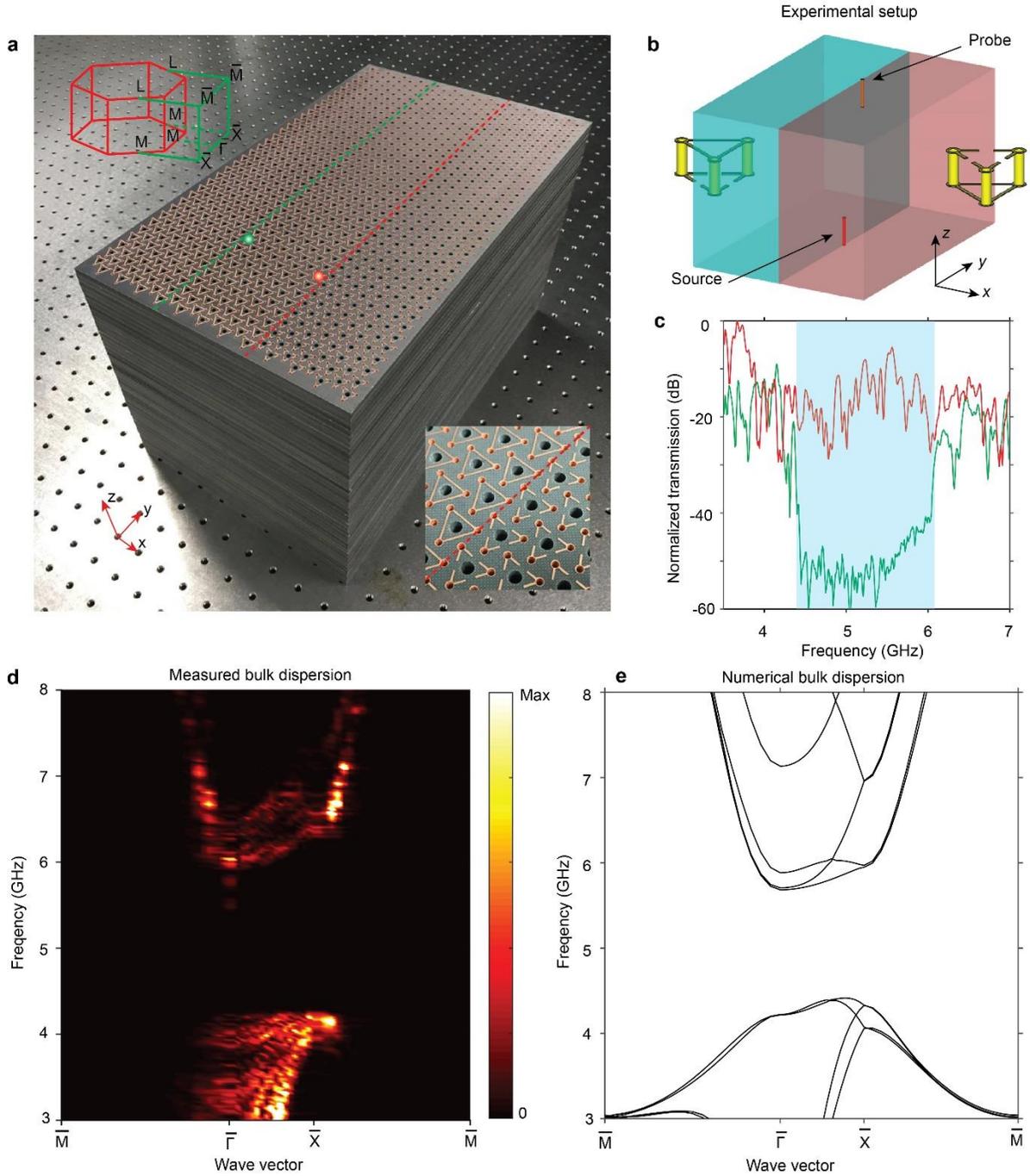

**Figure 2. Sample, experimental setup, and measured bulk dispersion of 3D PTI.** (a) Photograph of the sample. The blue and red lines indicate the planes used in measuring the bulk and surface states, respectively. Left top inset: projected Brillouin zone for the domain wall. Right bottom inset: photograph of the unit cells near the domain wall. (b) Experimental setup. The source is positioned on the bottom center of the surface, while the probe sweeps the selected plane hole-by-hole. (c) Measured transmissions at the probe point located a depth of 8 unit cells beneath the red (green) dot. The region highlighted in light blue represents the bulk topological bandgap. (d)-(e) Measured and simulated band diagrams of the 3D photonic TI, projected on the measurement plane. The color bar in (d) measures the electric energy density.

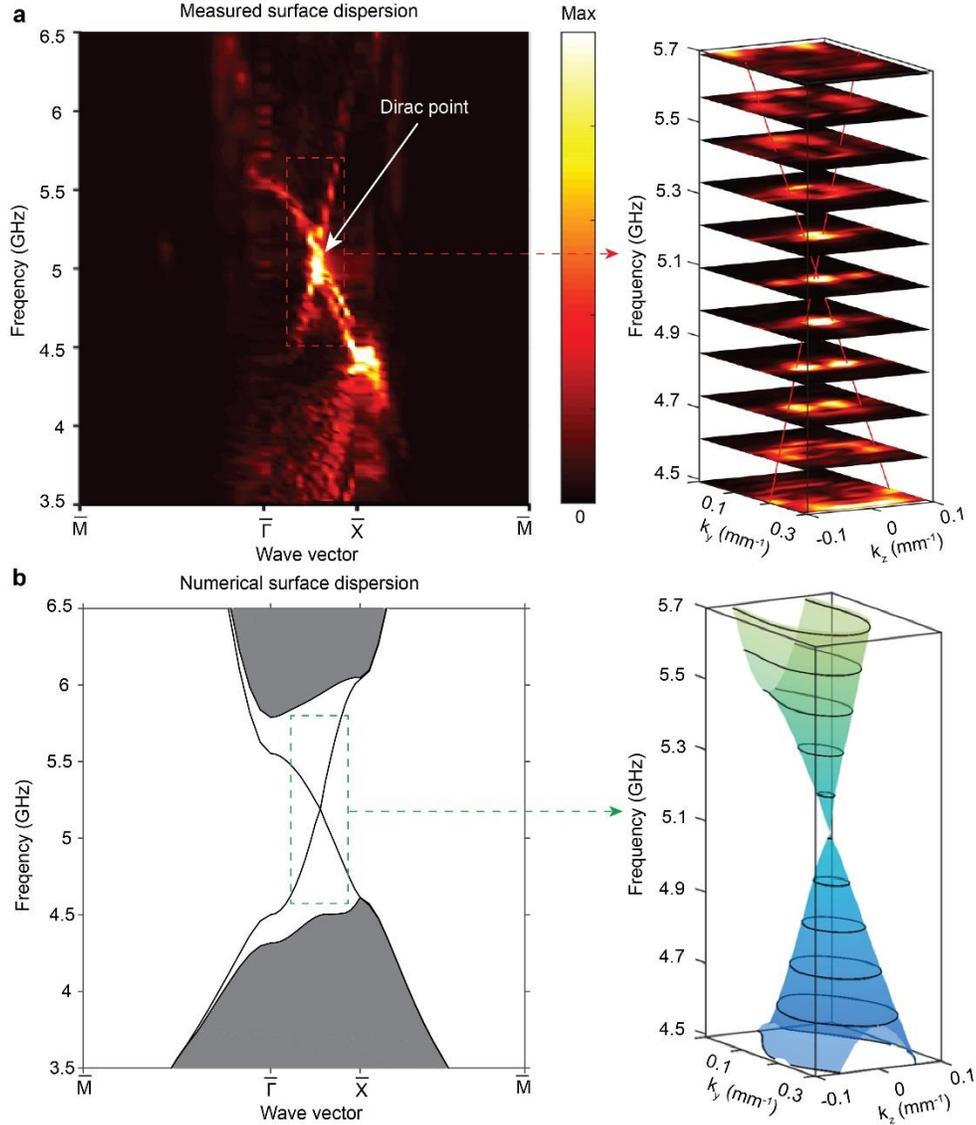

**Figure 3. Experimental observation of gapless conical Dirac-like topological surface states of the 3D photonic TI.** (a)-(b) Measured and calculated band diagrams for the topological surface states. Insets: isofrequency contours of the topological surface states at different frequencies. Red lines in (a) are guides to the eye that indicate the shape of the cones that intersect at the Dirac point. Note that there is another Dirac point near the projection of K′, not shown here. The color bar in (a) measures the electric energy density.

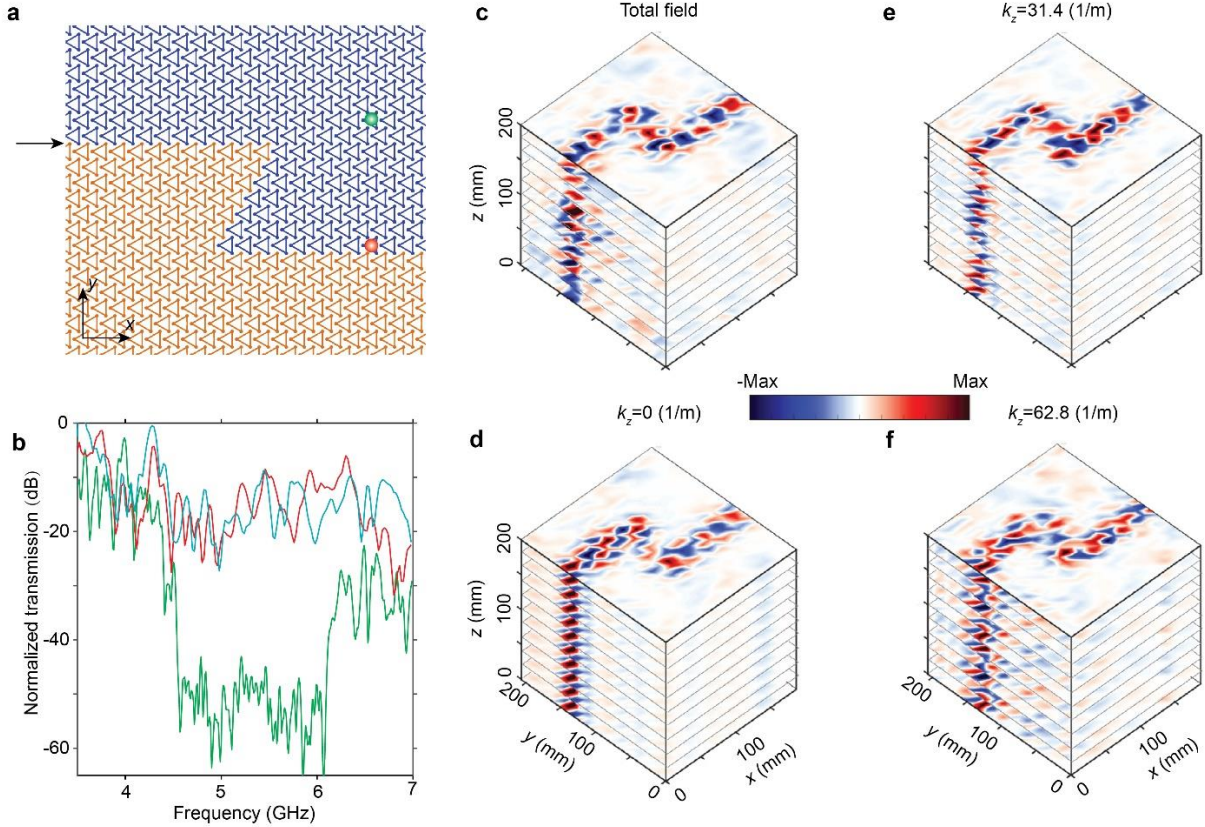

**Figure 4. Experimental demonstration of the robustness of photonic topological surface states.** (a) Cross-sectional schematic of the fabricated sharply twisted domain wall. The orange (blue) triangles in the upper (lower) domain represent the unit cells of SRRs with downward (upward) openings. A point-like source is located at the *x-y* position indicated by the black arrow, with *z* position at the middle of the sample. The red (green) dot indicates the *x-y* position of the detector used to measure the surface state transmission (bulk transmission). (b) Measured transmissions for surface states along twisted/straight domain wall (red/blue line) and bulk states (green line), respectively. The region highlighted in light blue represents the bulk topological bandgap. (c) Measured total field distribution in the *x-y* planes for different *z*, at 4.7 GHz. (d)-(f) Measured field distributions in the *x-y* planes with different *z* and different values of $k_z$, at 4.7 GHz.